\documentclass{nature_mod}

\usepackage[numbers, sort&compress]{natbib}
\usepackage{url}
\usepackage{epsfig}
\usepackage{graphicx}
\usepackage{xcolor}
\usepackage{caption}
\usepackage[colorlinks=true, linkcolor=blue, citecolor=blue, urlcolor=blue]{hyperref}
\usepackage{lineno}
\usepackage{aas_macros}
\usepackage{threeparttable}
\usepackage{multibib}
\newcites{methods}{Methods References}

\usepackage{amssymb,amsmath,caption}

\newcommand{\jcap}{J. Cosmol. Astropart. Phys.}

\long\def\symbolfootnote[#1]#2{\begingroup%
\def\thefootnote{\fnsymbol{footnote}}\footnote[#1]{#2}\endgroup} 

\title{Detecting Extragalactic Axion-like Dark Matter with Polarization Measurements of Fast Radio Bursts}

\author{Bao Wang$^{1,2}$\thanks{These authors contributed equally.},
	Xuan Yang$^{1,2 \ast}$,
	Jun-Jie Wei$^{1,2}$\thanks{Email: jjwei@pmo.ac.cn},
	Song-Bo Zhang$^{1}$\thanks{Email: sbzhang@pmo.ac.cn},
	Xue-Feng Wu$^{1,2}$\thanks{Email: xfwu@pmo.ac.cn}}

\begin{document}
\maketitle

\begin{affiliations}
 \item Purple Mountain Observatory, Chinese Academy of Sciences, Nanjing 210023, China
 \item School of Astronomy and Space Sciences, University of Science and Technology of China, Hefei 230026, China
\end{affiliations}

\bigskip

\begin{abstract}
Axions or axion-like particles (ALPs) are one of the promising dark matter (DM) candidates.
A prevalent method to detect axion-like DM is to seek periodic oscillation in the polarization angles (PAs)
of linearly polarized light emitted from astrophysical sources. In this work, we use the time-resolved
polarization measurements of the hyperactive repeating fast radio burst, FRB 20220912A, detected by
the Five-hundred-meter Aperture Spherical radio Telescope (FAST) to search for extragalactic axion-like DM. 
Given a DM density profile of FRB 20220912A's host, we obtain upper  
limits on the ALP-photon coupling constant of $g_{a \gamma}<(3.4 \times 10^{-11}-1.9\times 10^{-9})\,\mathrm{GeV}^{-1}$
for the ALP masses $m_a \sim (1.4\times10^{-21}-5.2\times10^{-20})\,\mathrm{eV}$. 
Persistent polarimetric observations with FAST would extend the constraints to lower masses. 
Although the $g_{a \gamma}$ constraints derived from FRBs are less competitive than those from 
other methods, FRBs offer an alternative way to detect axion-like DM on extragalactic distance scales, 
complementary to galactic DM probes.
\end{abstract}

\clearpage

\section*{Introduction}\label{sec:intro}

One of the most important tasks that remain unresolved in modern physics is the detection of dark matter
particles (DM, hereafter referring to dark matter rather than dispersion measure), and numerous candidates of DM have been proposed. Axion-like particles (ALPs), a type of ultralight
bosons, have emerged as the most prevailing candidates in the search for DM \cite{Preskill1983, Abbott1983, Dine1983, Khlopov1999, Marsh2016, Choi2021}.
Thanks to the special properties of ultralight scalar particles ($m_a \sim 10^{-22} \;{\rm eV}$), they can provide
a natural solution to the challenges encountered in small-scale structures of the Universe \cite{Hu2000, Hui2017, Hui2021}.
A lot of strategies for finding axion-like DM have been explored, including photons from axion conversion
\cite{Horns2013, Payez2015, Anastassopoulos2017, Du2018}, nuclear magnetic resonance \cite{Graham2013, Budker2014},
periodic oscillations of linearly polarized light \cite{Ivanov2019, Chen2020, Liu2020,  Yuan2021, Gan2024}, and
other terrestrial or astronomical experiments \cite{Dessert2020, Meyer2020}.

Among the various axion-like DM detection methods, detecting periodic oscillations of polarized light has been
regarded as a promising approach in astrophysics \cite{Fedderke2019, Fujita2019, Caputo2019, Ivanov2019, Chen2020, Liu2020, Yuan2021, Castillo2022, Liu2023b, Gan2024}.
When light propagates through the ALP field, photons would interact with ALPs, and the interaction term is
$\mathcal{L}_{a \gamma} =\frac{1}{4}g_{a\gamma} a F_{\mu \nu} \tilde{F}^{\mu \nu}$,
where $g_{a\gamma}$ represents the coupling strength between the axion field ($a$) and the electromagnetic field
($F_{\mu \nu}$). The interaction leads to modifications in the dispersion relations \cite{Carroll1990, Harari1992}.
The left- and right-handed circular polarization modes of light experience opposite corrections due to their
different dispersion relations. This effect is known as cosmic birefringence, resulting in changes in
the polarization angles (PAs) of the light. Therefore, in the presence of an ALP field, if the light is linearly polarized,
its PAs will oscillate with the ALP field, with an amplitude proportional to $g_{a\gamma}$.

\begin{figure*}
	\centering
	\includegraphics[width=0.98\textwidth]{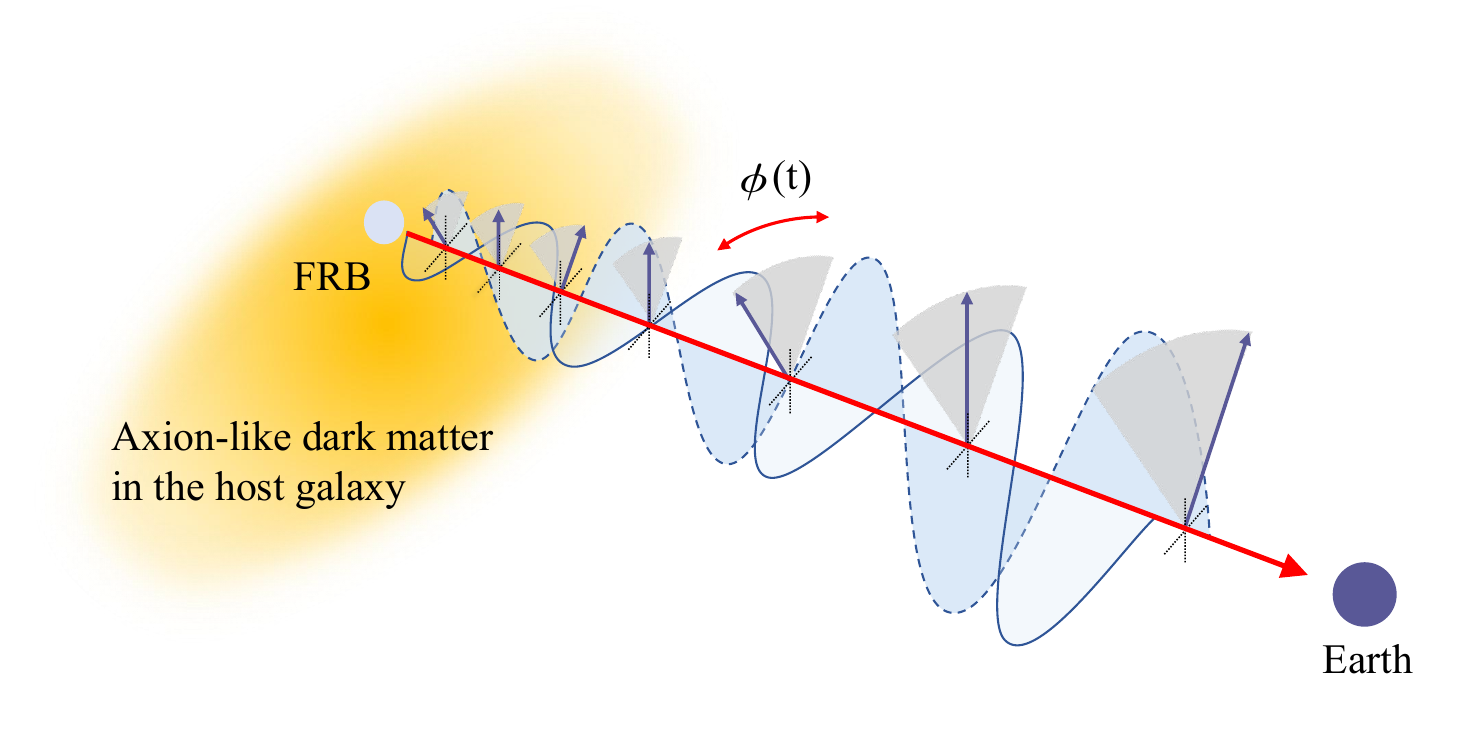}
	\caption{
		{\bf Illustration of detecting extragalactic axion-like dark matter (DM) with polarization measurements of FRBs.} 
		The interaction between photons and ALPs leads to modifications in the dispersion relations, 
        resulting in a difference in phase velocity between the two modes.
		This phenomenon, known as cosmic birefringence, causes changes in the polarization angles (PAs) of the light.
            If axion-like DM is distributed around the FRB 20220912A's host galaxy, ALP-induced PA oscillations ($\phi(t)$) can emerge.
		The repeating FRBs enable us to monitor their polarization properties long-term to detect axion-like DM on extragalactic distance scales.
	} 
	\label{Figure 1}
\end{figure*}

Fast radio bursts (FRBs) are brief and intense radio transients that originate from cosmological distances \cite{Lorimer2007, Thornton2013, Spitler2016, Xiao2021, 2022A&ARv..30....2P, Chime2022, Zhang2023b}.
FRBs have been powerful astrophysical laboratories for studying
cosmology \cite{Wei2015, Wu2016, Yang2016, Walters2018, Macquart2020, Wei2021, Beniamini2021, Yang2022, James2022, Liu2023a, Wang2023, Zhang2023a}
and also have the potential to play an important role in the detection of DM \cite{Caputo2019, Landim2020, Sammons2020, 2016PhRvL.117i1301M, Wang2018, Liao2020, Ho2023, Krochek2022, Gao2023}.
To date, no studies have utilized real polarization observations of FRBs to constrain axion-like DM directly.
The necessary conditions for FRBs to serve as axion-like DM probes include: (i) active repeating bursts, (ii) nonmagneto-ionic local environments with stable PAs, (iii) highly linear polarization, and (iv) precise localizations within host galaxies.
The repetition pattern of FRBs enables us to monitor their polarization properties long-term
to detect axion-like DM on extragalactic distance scales, complementary to other galactic DM probes.
The schematic illustration is shown in Fig.~\ref{Figure 1}.
One intriguing sample for such a study is FRB 20220912A, an active repeating source with highly linear polarization \cite{Ravi2023, Zhang2023}, and its local environment is nonmagneto-ionic \cite{Feng2024}. 
Its long-term polarization observations from the Five-hundred-meter Aperture Spherical radio Telescope (FAST) \cite{Zhang2023} provide a remarkable opportunity
to detect extragalactic axion-like DM through searching for a periodic oscillation in the PAs.

In this work, we constrain the ALP-photon coupling constant $g_{a\gamma}$ using polarization data of FRBs.
We analyze the polarization angle variations of linearly polarized emission from FRB 20220912A
observed by FAST. All currently available observations from October 28th, 2022 to December 5th, 2022 are adopted
for our study. The observational time coverage of $\sim 38$ days with a cadence of $\sim 1$ day is sensitive to
ALPs with mass ranging from $1.4\times 10^{-21}$ eV to $5.2\times 10^{-20}$ eV. By estimating the periodic variations in
linear PAs, we can place upper limits on $g_{a\gamma}$ directly. Finally, we also predict further constraints
from continued polarization observations of FRB 20220912A in the future.

\section*{Results}

\subsubsection*{The Polarization Angles detected by FAST}

After the initial discovery of FRB 20220912A, subsequent observations from multiple telescopes have
consistently detected a large number of bursts from this specific source \cite{Bhusare2022,Fedorova2022,
Herrmann2022,Kirsten2022,Ould2022,Pelliciari2022,Perera2022,Rajwade2022,Sheikh2022,Yu2022,Zhang2023,Feng2024}.
It is noteworthy that FRB 20220912A was monitored by FAST for a period of several dozen days,
during which a total of 1076 bursts were recorded \cite{Zhang2023}. Most of these bursts exhibit
nearly 100\% linear polarization. The rotation measure (RM) of FRB 20220912A is very close to 0
and did not show any variation during the FAST observation period, indicating that FRB 20220912A
is located in a likely nonmagneto-ionic local environment \cite{Zhang2023, Feng2024}. The non-variable
RM also means that the PAs of FRB 20220912A are relatively stable.

\begin{figure*}
  \centering
  \includegraphics[width=0.99\textwidth]{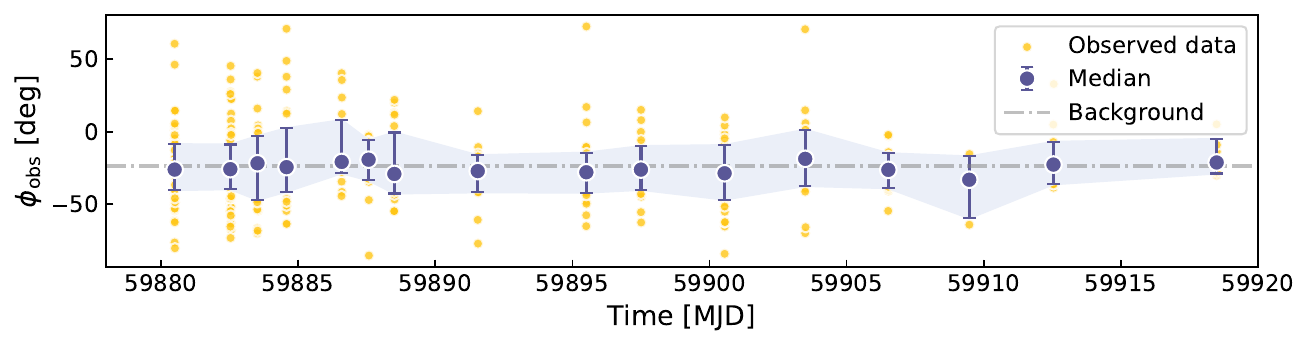}
  \caption{
  {\bf Linear polarization measurements of the bursts from FRB 20220912A detected by FAST.} 
  The observed polarization angle (PA, $\phi_{\rm obs}$) of each burst is depicted by a yellow point, the purple points represent
  the daily medians, and the light purple shaded area encompasses the $1\sigma$ confidence range.
  The dashed line represents the mean background $ \langle  \phi_{\rm bkg}  \rangle = -24.98^{\circ}$.
  } 
 \label{Figure 2}
\end{figure*}

In our analysis, the PAs of FRB 20220912A are processed from the raw data of FAST. 
The detailed data processing can be found in the subsection \hyperlink{Data}{The PA Data Analysis} of Methods.
Fig.~\ref{Figure 2} displays PA measurements of FRB 20220912A as a function of time (MJD 59880 to MJD 59918), and the median value of the PAs is calculated for each day.
As illustrated in this plot, the PAs exhibit considerable variation within a day, but they are relatively stable on the monthly timescale during the observational period.
Because of the timing of the observations, there is a concentration of data for a single day within narrow time intervals, which results in poor data continuity from one day to the next.
Therefore, our analysis focuses on PA variations with a minimum oscillation period of one day.
In this case, the minimum time interval is 1 day, while the total observational time is 38 days.
According to the theory presented in the subsection \hyperlink{Coupling}{ALP-photon Coupling} of Methods,
the ALP mass $m_a$ can be determined through $m_a = 2\pi(1+z)/T'$. It is clear from this formula that the lower and upper limits of
the mass $m_a$ depend on the total observational time ($T'=38$ days) and the minimum time interval ($T'=1$ day), respectively.
That is, the sensitive ALP mass $m_a$ falls within the range of $1.4\times 10^{-21}$ eV to $5.2\times 10^{-20}$ eV.

\subsubsection*{Search for ALP-induced Oscillations}

As described in the subsection \hyperlink{Coupling}{ALP-photon Coupling} of Methods, when a linearly polarized light propagates in an external ALP field,
the corresponding PAs would have a time-dependent change due to the ALP-photon
coupling effect. By analyzing the periodic variations of PAs, we thus can constrain the ALP-photon
coupling strength, thereby facilitating the potential detection of axion-like DM. 

\begin{figure*}
  \centering
  \includegraphics[width=0.9\textwidth]{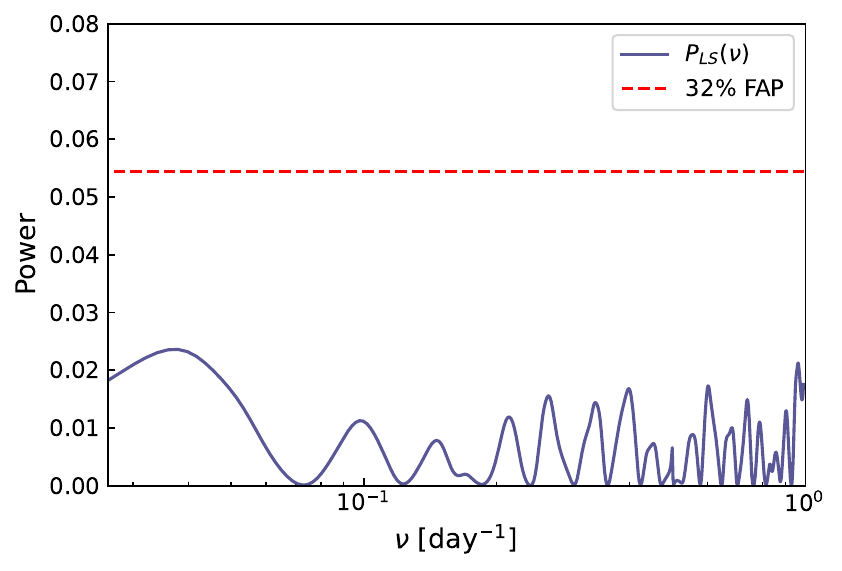}
  \caption{
  {\bf The LS periodogram for the time series of polarization measurements for FRB 20220912A.} 
  The purple line represents the power spectrum ($P_{\rm LS}(\nu)$) of the LS periodogram.
  The red dashed line marks the False Alarm Probability (FAP) threshold at 32\%. 
  The signals with $P_{\rm LS}(\nu)$ values below this line are not considered to oscillate periodically at the 1$\sigma$ confidence level.
  } 
 \label{Figure 3}
\end{figure*}

The  Lomb-Scargle (LS) Periodogram is a general tool for searching periodic signals, and has also been applied to axion detection \cite{Ivanov2019, Castillo2022}.
The significance of the peak values in the LS Periodogram can be assessed by the False Alarm Probability (FAP).
The LS periodogram and FAP can be calculated using the python package \emph{\sc Astropy}, and the results from the time series of polarization measurements for FRB 20220912A with ionospheric corrections are shown in Fig.~\ref{Figure 3}.
More details can be found in the subsection \hyperlink{LSP}{The Lomb-Scargle Periodogram} of Methods.
We can find that all values are much lower than the 32\% FAP line.
Consequently, there is insufficient evidence to support ALP-induced periodic oscillations present in the PAs of FRB 20220912A based on current observations.
Therefore, we can only determine an upper limit of the ALP-photon coupling constant $g_{a \gamma}$ using an alternative method.

\subsubsection*{The Resulting Constraints and Comparisons}
	
We employ two analysis methods to constrain the ALP-photon coupling constant: the standard deviation (SD) method and the LS Periodogram-Monte Carlo method \cite{Lomb1976, Scargle1982, VanderPlas2018}.
The former is cruder but more convenient for estimation.
Further details of the two methods can be found in the subsections \hyperlink{SD of PA}{The Standard Deviation of PAs} and \hyperlink{MC}{LS Periodogram-Monte Carlo Method} of Methods, respectively.

\begin{figure*}
  \centering
  \includegraphics[width=0.99\textwidth]{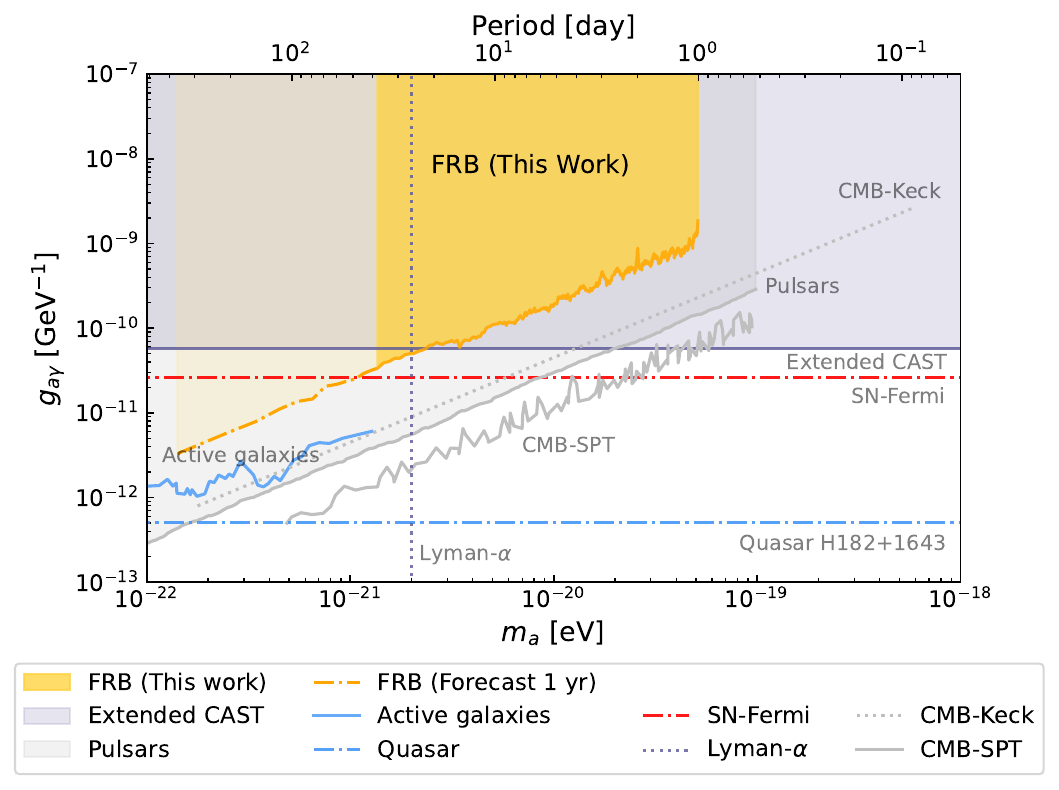}
  \caption{
  {\bf The resulting constraints on the ALP-photon coupling constant $g_{a \gamma}$ for different ALP mass $m_a$, obtained from the polarization measurements of FRB 20220912A.} 
  	The yellow shaded area corresponds to the upper limits of $g_{a \gamma}$ derived from the LS Periodogram-Monte Carlo method.
    The yellow dot-dashed line represents future constraints from
  continued polarization observations of FRB 20220912A for up to one year.
  Other 95\% CL upper limits from different
  astrophysical sources are also displayed for comparison, including the VLBA polarization observations of jets from active galaxies (blue solid line) \cite{Ivanov2019},
  the Chandra observation of the quasar H1821+643 (blue dot-dashed line) \cite{Sisk2022},
  the Extended CAST experiment (purple shaded area) \cite{CAST2024}, 
   the polarized light of pulsar from the Parkes Pulsar Timing Array (PPTA) project (gray shaded area) \cite{Castillo2022}, 
   the Fermi-LAT observation of supernovae (red dot-dashed line) \cite{Meyer2020},
   the observations of the cosmic microwave background (CMB) from BICEP/Keck (gray dotted line) \cite{Ade2022} and SPT-3G (gray solid line) \cite{Ferguson2022},
   and the mass constraint from the Lyman-$\alpha$ forest data (vertical dotted line) \cite{Irsic2017}.
  For a more complete summary, please refer to this link \cite{footnote}.
  } 
 \label{Figure 4}
\end{figure*}

For the estimation from the SD method, we obtain upper limits of 
$g_{a \gamma}<(2.7 \times 10^{-11}-1.0\times 10^{-9})\,\mathrm{GeV}^{-1}$ for the ALP masses 
$m_a \sim (1.4\times10^{-21}-5.2\times10^{-20})$ eV. 
For the LS Periodogram-Monte Carlo method, the obtained upper
limits are $g_{a \gamma}<(3.4 \times 10^{-11}-1.9\times 10^{-9})\,\mathrm{GeV}^{-1}$ for 
the same ALP mass range, which is consistent with the results of the first method.
The resulting constraints on $g_{a \gamma}$ from the LS Periodogram-Monte Carlo method are shown in Fig.~\ref{Figure 4},
along with other 95\% CL upper limits from different astrophysical sources. 
Finally, we also forecast a future constraint.
If the polarization observations of FRB 20220912A last up to one year, the limit obtained from the LS Periodogram-Monte Carlo method would extend to lower ALP masses, yielding $g_{a \gamma}<3.3 \times 10^{-12}\,\mathrm{GeV}^{-1}$ for an ALP mass $m_a \sim 1.4\times 10^{-22}$ eV.
	
Recently, Gao et al. \cite{Gao2023} proposed gravitationally lensed FRBs as probes for hunting Galactic axion DM, predicting that the ALP-photon coupling could be constrained to be
$g_{a \gamma}<7.3 \times 10^{-11}\,\mathrm{GeV}^{-1}$ for an axion mass $m_a \sim 10^{-21}$ eV.
This forecast limit is similar to our result of $g_{a \gamma}<2.1 \times 10^{-11}\,\mathrm{GeV}^{-1}$
for the same axion mass from real FRB polarization observations.
Furthermore, as shown in Fig.~\ref{Figure 4}, our result is
slightly better than the constraint from the extended CERN Axion Solar Telescope (CAST) ($g_{a \gamma}<5.8\times 10^{-11}\,\mathrm{GeV}^{-1}$)
\cite{CAST2024} and comparable with the constraint from the supernovae observed by
the Fermi Large Area Telescope (LAT) ($g_{a \gamma}<2.6\times 10^{-11}\,\mathrm{GeV}^{-1}$) \cite{Meyer2020}.
In contrast to the galactic probes, such as pulsars \cite{Caputo2019, Liu2020,  Castillo2022}, black hole \cite{Yuan2021, Gan2024}, and protoplanetary disk \cite{Fujita2019}, our method can detect ALPs
on kiloparsec scales, which highlights the potential of FRBs for detecting extragalactic axion-like DM.

\section*{Discussion} 
In this work, we attempted to detect extragalactic ALPs in the host galaxies of the hyperactive repeating FRBs. 
Thanks to the nonmagneto-ionic local environment, highly linear polarization, and relatively stable PAs within the observation period, 
FRB 20220912A is hitherto the most suitable repeating source for carrying out such a study. 
Note that the distance of FRBs does not provide an advantage in detecting axion-like DM. This is because the PA shift depends on the axion field at the initial and final positions, independent of the propagation distance (see Equation (\ref{eq:PA1}) in Methods).
With the state-of-the-art PA observations of the bursts from the repeating source FRB 20220912A taken by FAST, we obtained upper limits on the ALP-photon
coupling constant of $g_{a \gamma}<(3.4 \times 10^{-11}-1.9\times 10^{-9})\,\mathrm{GeV}^{-1}$ for
the ALP masses $m_a \sim (1.4\times10^{-21}-5.2\times10^{-20})$ eV. If the polarization observations of
FRB 20220912A are expected to last for one year, the $g_{a \gamma}$ limit would be $g_{a \gamma}<3.3 \times 10^{-12}\,\mathrm{GeV}^{-1}$ for an ALP mass $m_a \sim 1.4\times10^{-22}$ eV.

Although the constraints on $g_{a \gamma}$ from FRBs are not as competitive as those from other sources, such as the active galaxies \cite{Ivanov2019}, the quasar H1821+643 and the cosmic microwave background \cite{Ade2022, Ferguson2022}, and the mass range of $m_a<2\times10^{-21}$ eV have been excluded by the Lyman-$\alpha$ Forest Data \cite{Irsic2017}, our attempt can serve as an alternative and complementary method.
FRBs have the advantage of being abundant in extragalactic systems.
Numerous extragalactic FRBs offer an alternative way to detect axion-like DM in various DM-rich extragalactic systems, thereby obtaining tighter constraints on $g_{a \gamma}$ that complement other galactic DM probes, such as pulsars.

In the era of Square Kilometre Array (SKA), a large sample of localized FRBs will significantly enhance their applications.
When numerous FRBs are combined, the constrained precision on the ALP-photon
coupling constant will statistically increase by a factor of $\sqrt{N}$.
According to Zhang et al. \cite{Zhang2023a}, $\mathcal{O}(10^5)$–$\mathcal{O}(10^6)$ FRBs can be detected by the mid-frequency array of the first phase of SKA (SKA1-MID) in a 10-year observation period.
To date, more than 800 FRBs have been detected \cite{Xu2023}, and among them, only one, FRB 20220912A, has been identified as a repeating FRB with a nonmagneto-ionic environment.
Assuming there are $N\sim 100$ (0.1\% of the total) FRBs like FRB 20220912A in SKA1-MID, the upper limit of the coupling constant would improve by approximately an order of magnitude.

It should be underlined here that the implicit assumption on our results is that the ALP mass exceeds the Hubble parameter in the appropriate epoch of the Universe.
In this condition, the ALP-induced PA oscillations can begin and lead to an evolution of energy density in the form of DM \cite{Marsh2016, Hui2021}.
If the ALP mass is so low that it approaches the Hubble constant, $m_a \sim 10^{-33}$ eV, it will be frozen and manifest as a form of dark energy rather than DM.

\section*{Methods}

\hypertarget{Coupling}{}
\subsubsection*{ALP-photon Coupling}
\label{sec:Coupling}

ALPs can be described as a pseudo-scalar field $a(x,t)$ with mass $m_a$, where $x$ is the spatial coordinates and $t$ is the time.
The ALP field can interact with the electromagnetic field, and its dynamics can be captured by the Lagrangian terms \cite{Wilczek1987}:
\begin{equation}\label{eq:Lagrangian}
	\mathcal{L} =-\frac{1}{4} F_{\mu \nu} F^{\mu \nu}+\frac{1}{2} \partial^{\mu} a \partial_{\mu} a-\frac{1}{2} m_{a}^{2} a^{2}+\frac{1}{4}g_{a\gamma} a F_{\mu \nu} \tilde{F}^{\mu \nu},
\end{equation}
where $F^{\mu \nu}$ denotes the electromagnetic field tensor,
$\tilde{F}^{\mu \nu}=\frac{1}{2}\epsilon^{\mu \nu \lambda \sigma} F_{\lambda \sigma}$ is the dual of $F^{\mu \nu}$, and
$g_{a\gamma}$ represents the ALP-photon coupling constant which characterizes the strength of interaction.
This coupling leads to a modification in the dispersion relation \cite{Ivanov2019}:
\begin{equation}
    \omega_{\pm} \simeq k \pm \frac{1}{2} g_{a \gamma} n^{\mu}\partial_{\mu}a,
\end{equation}
where $n^{\mu}$ is null tangent vector of light, $k$ is the wave vector,
and the frequency $\omega_{\pm}$ corresponds to two circular polarization states. 
The natural unit system $\hbar$=c=1 is employed here.
When two vertically polarized
electromagnetic waves of these two states propagate, a phase shift occurs between them due to the disparity
in their phase velocities. This phase shift leads to the rotation of the polarization plane, known as
cosmic birefringence. Specifically, the frequency difference between the two polarization components is
$\Delta \omega = \omega_{+}-\omega_{-}= g_{a \gamma}  n^{\mu}\partial_{\mu}a$. If waves emitted from the source
at position $x_1$ at time $t_1$ are received by an observer at position $x_2$ and time $t_2$, the ALP-induced
PA shift is then expressed as
\begin{equation}\label{eq:PA1}
\begin{split}
    \phi & = \frac{1}{2}\int_{C}\Delta \omega  {\rm d} s 
    = \frac{1}{2} g_{a \gamma} \int_{x_1}^{x_2} \partial_{\mu}a  {\rm d} x  \\
    & = \frac{1}{2}  g_{a \gamma} \left[ a(x_2, t_2) - a(x_1, t_1) \right],
\end{split}
\end{equation}
where $C$ is the propagation path of waves.
From Equation (\ref{eq:PA1}), it is evident that the PA shift $\phi$ is determined by the time-dependent axion field $a(x,t)$ at the initial and final positions, since it arises from the path integral of the axion field gradient ($\partial_{\mu}a$).
The equation of motion for the ALP field is given by the Klein-Gordon equation.
When we neglect the friction term, the solution simplifies and exhibits an oscillating form:
\begin{equation}
	a(x, t) = a_0(x) \sin \left(m_a t+\delta\right),
\end{equation}
where $\delta$ is the position-dependent phase. $a_0(x)$ is the amplitude that relates to the energy density
of the ALP field $\rho$ (or equivalently the energy density of DM, if the dominant DM is assumed to be made up of
ALPs), i.e., $\rho=\frac{1}{2} m_a^2 \alpha^{-2}  a_0^2$, 
where $\alpha$ is a random nonnegative variable following the Rayleigh distribution $f(\alpha)=\alpha\exp{( -\alpha^2/2)}$ \cite{Foster2018}.
When the observed time scale is much smaller than the coherence time scale, it becomes essential to consider this stochastic nature of the connection between the amplitude $a_0$ and the energy density $\rho$ \cite{Foster2018, Centers2021}.
The oscillation period of the ALP field is given by $T=2 \pi/ m_a$,
which depends on the ALP mass. If the energy density of the ALP field at the observer is much lower than
the one at the source (i.e., $a(x_2, t_2) \ll a(x_1, t_1)$), Equation~(\ref{eq:PA1}) can be converted to
an oscillatory expression,
\begin{equation}\label{eq:PA2}
	\phi (t) = - \frac{\sqrt{2}}{2} g_{a \gamma}\,  m_a^{-1}
	\rho^{\frac{1}{2}} \alpha
	\sin \left(2 \pi \frac{t}{T^{\prime}}+\delta \right),
\end{equation}
where $T'=T(1+z)$ is the observed period on Earth, taking into account cosmic expansion.
Equation~(\ref{eq:PA2}) describes that the PAs have the periodic oscillation characteristic
when linearly polarized waves are coupled with ALPs.

\subsubsection*{DM Density Profile of The Host Galaxy}
\label{sec:cal}

Outside the solitonic cores of galaxies, the DM density distribution $\rho (r)$ can be approximately described by the generalized
Navarro-Frenk-White (NFW) profile \cite{Zhao1996}:
\begin{equation}\label{eq:darkmatter}
	\rho(r)=\frac{\rho_{0}}{\left(r / r_{\mathrm{s}}\right)^{\beta}\left(1+r / r_{\mathrm{s}}\right)^{3-\beta}} ,
\end{equation}
where $r$ is the distance from the galaxy center, $\rho_0$ is the characteristic density, $r_s$ is the scale radius,
and $\beta$ is the power-law index. Also, $\rho(r) \propto  r^{-\beta}$ when $r \ll r_s$ and $\rho(r) \propto r^{\beta-3}$
when $r \gg r_s$. For the case of $\beta=1$, Equation~(\ref{eq:darkmatter}) is reduced to the original NFW profile \cite{Navarro1996}.
In principle, these coefficients ($\rho_0$, $r_s$, and $\beta$) can be determined by fitting the rotation curves of galaxies.
The physical origins of FRBs are still unknown, but some of them have been localized in extragalactic host galaxies.
Once we have enough observational information about the FRB host galaxy, we can estimate the DM density $\rho$
in the vicinity of the FRB source.

The Deep Synoptic Array localized the repeater FRB 20220912A to a host galaxy, PSO J347.2702+48.7066,
at redshift $z = 0.0771$ \cite{Ravi2023}. The host galaxy has a stellar mass of approximately
$10^{10}\,M_{\odot}$, a star-formation rate of $\gtrsim 0.1\,M_{\odot} \, {\rm yr^{-1}}$, and an effective
radius of $2.2\,{\rm kpc}$. Gordon et al. \cite{2023ApJ...954...80G} compared the optical host luminosities of repeating
and nonrepeating FRBs across redshift, and defined a demarcation at luminosity $10^{9}\,L_{\odot}$
below which a host can be classified as a dwarf galaxy. FRB 20220912A sits just above the borderline
at $\approx 1.1\times10^{9}\,L_{\odot}$, suggesting that
its host may be a dwarf galaxy \cite{2023ApJ...954...80G}.

Since they have higher fractions of DM compared to more massive systems, dwarf galaxies are deemed as
ideal systems to probe the DM density profile \cite{Cooke2022}. However, we currently lack rotation curve
observations of the host of FRB 20220912A to investigate its DM density distribution. Here we use
the DM density profile of a dwarf galaxy, NGC 4451 (with a similar stellar mass of $\sim 10^{10}\,M_{\odot}$
and a similar radius of $\sim 2.2\,{\rm kpc}$ \cite{Cooke2022}), as a reference. Note that the differences in
DM density profiles between NGC 4451 and FRB 20220912A's host have negligible effects within the precision
range of our study. Based on the stellar rotation curve, Cooke et al. \cite{Cooke2022} determined the coefficients of the
generalized NFW profile (Equation~(\ref{eq:darkmatter})) for NGC 4451, i.e., $\rho_0=0.41\,M_{\odot}\,{\rm pc^{-3}}$,
$r_s=2.2\,{\rm kpc}$, and $\beta=0$. Furthermore, a recent milliarcsecond localization of FRB 20220912A
shows that its transverse offset from the host galaxy center is $r\approx0.8$ kpc \cite{hewitt2023}.
With this information, an estimate of the DM density at the location of FRB 20220912A from
Equation~(\ref{eq:darkmatter}) is $\rho \sim 0.16 \,M_{\odot} \, {\rm pc^{-3}}$. This value is much larger than
the DM density near our Earth, which is $\sim 0.01 \,M_{\odot} \, {\rm pc^{-3}} $ estimated by
the Galactic NFW profile \cite{Nesti2013}.

\hypertarget{Data}{}
\subsubsection*{The PA Data Analysis}

The PAs of FRB 20220912A were derived from the raw data of FAST. The central frequency, bandwidth, number of frequency channels, and sampling time for the raw data were 1.25 GHz, 0.5 GHz, 4096, and 49.152 $\rm \mu s$, respectively. We used the GPU-accelerated transient search software \emph{\sc HEIMDALL}
and processed the data on FAST's high-performance computer facilities. 
A dispersion measure range of 200 to 250 $\rm pc\,cm^{-3}$ was searched, with a signal-to-noise ratio threshold of 6.5 and a maximum boxcar of 512. 
After determining the dispersion measures, the de-dispersed polarization data were calibrated using the \emph{\sc psrchive} software package with correction for differential gain and phase between the receivers achieved through the injection of a noise diode signal before each observation.
The rotation of the telescope and the variation of the receiver across the days were calibrated through \emph{\sc pac}. 
The RM was searched from $-2000$ to $2000$ $\rm rad\,m^{-2}$ in steps of 1 $\rm rad\,m^{-2}$ using the \emph{\sc rmfit} program \cite{Straten12}. Ionospheric RMs in the direction of FRB 20220912A at each burst's arrival time were computed using FRion package \cite{Van23}. The ionosphere model is sourced from the International GNSS Service (IGS), which provides ionosphere vertical total electron content (TEC) maps daily. FRion downloads these TEC maps from NASA CDDIS archive.
After correcting the data with the best-fitted RMs, we derive the PAs of the linearly polarized component. During a total of 9.2 hours of observations between October 28th, 2022 and December 5th, 2022 (corresponding to MJD 59880 and MJD 59918), we obtain 674 bursts with RM measurements. 
The PA data of these bursts are available in Supplementary Data 1.

\hypertarget{SD of PA}{}
\subsubsection*{The Standard Deviation of PAs}

Variations in PAs of FRBs are complex and puzzling \cite{Cho2020,Luo2020,Xu2022}. The prevailing understanding
is that these variations are mainly attributed to the significant fluctuations in the magnetic fields surrounding
FRB sources. However, if the axion-like DM exists in the host galaxy and envelops the FRB sources, the observed PA
($\phi_{\rm obs}$) is expected to be composed of two components: one from the PA contribution of the astrophysical
background (e.g., the magnetic field), $\phi_{\rm bkg}$, and the other one is the ALP-induced PA shift,
$\phi (t)$, i.e.,
\begin{equation}\label{eq:PAtot}
	\phi_{\rm obs} = \phi_{\rm bkg} + \phi (t) 
	\simeq \langle  \phi_{\rm bkg}  \rangle  + \Delta \phi_{\rm bkg} + \phi (t),
\end{equation}
where $ \langle \phi_{\rm bkg} \rangle $ represents the mean value of the PA caused by the background
magnetic field and $\Delta \phi_{\rm bkg}$ corresponds to the PA fluctuation arising from the magnetic field changes.
Since $\Delta \phi_{\rm bkg}$ is unpredictable, we simply assume that the magnetic field is time-invariant, which 
means that the observed PA fluctuations are attributed to
the ALP-photon coupling effect, i.e., $\phi_{\rm obs}=  \langle  \phi_{\rm bkg}   \rangle + \phi (t)$.
Actually, the fluctuations caused by time-varying magnetic fields are quite real, which means that ignoring the contribution
from $\Delta \phi_{\rm bkg}$
will conduct conservative upper limits on the ALP-photon coupling constant $g_{a \gamma}$ for different ALP masses, except in a case of coincidences where $\phi (t)$ and $\Delta \phi_{\rm bkg}$ cancel out in opposite phases.
	
Given the randomness of the value of the phase $\delta$ (see Equation (\ref{eq:PA2})), we use the SD of ALP-induced PA shift $\phi (t)$ to characterize its oscillation amplitude \cite{Liu2020}. This yields
\begin{equation}\label{eq:PAstd}
		\begin{split}
			\Delta \phi  \equiv \sqrt{ \left \langle \phi^2 (t)  \right \rangle} 
			& \simeq 1.96^{\circ} \left( \frac{\rho}{0.16\, M_{\odot} \, {\rm pc^{-3}}}\right)^{\frac{1}{2}} \\
			& \times \left(\frac{m_a}{10^{-21}\, \mathrm{eV}}\right)^{-1}
			\left( \frac{g_{a \gamma}}{10^{-11}\, \mathrm{GeV}^{-1}}\right).
\end{split}
\end{equation}
The mean value of the observed daily median PAs is $\langle \phi_{\rm Med} \rangle=-24.98 \pm 3.83 \, ^{\circ}$.
Here the mean $-24.98^{\circ}$ is regarded as the mean background $\langle\phi_{\rm bkg} \rangle$,
and the $1\sigma$ SD $3.83^{\circ}$ is regarded as $\Delta \phi$. From Equation~(\ref{eq:PAstd}),
we can see that $\Delta \phi$ is proportional to $g_{a \gamma}$ for a given ALP mass $m_a$. With the ALP mass
ranging from $1.4\times 10^{-21}$ eV to $5.2\times 10^{-20}$ eV, the corresponding upper limits on $g_{a \gamma}$
can be obtained as $g_{a \gamma}<(2.7 \times 10^{-11}-1.0\times 10^{-9})\,\mathrm{GeV}^{-1}$.

\hypertarget{LSP}{}
\subsubsection*{Lomb-Scargle Periodogram}

The LS periodogram is a commonly used technique to identify the periodic signals in time series \cite{Lomb1976, Scargle1982}.
It is widely applied in astronomy \cite{Zechmeister2009, VanderPlas2018}, and has also been employed in axion search \cite{Ivanov2019, Castillo2022}.
The LS Periodogram involves the computation of the power spectrum $P_{\rm LS}(\nu)$, which is associated with the probability of a periodic signal at frequencies $\nu$. A higher $P_{\rm LS}(\nu)$ value indicates a greater probability of periodicity.
We consider a time series $(y_i,\, t_i)$ with SD $\sigma_i$ of length $N$ ($i=1,...,N$), and then the required symbols are defined as follows: \cite{Zechmeister2009}
	\begin{equation}
		\begin{split}
			Y_{Y} & =\sum_{i=1}^{N} w_{i} y_{i}^{2}-Y \cdot Y , \qquad \qquad  \!
			Y_{C} =\sum_{i=1}^{N} w_{i} y_{i} c_{i}-Y \cdot C , \\
			Y_{S} & =\sum_{i=1}^{N} w_{i} y_{i} s_{i}-Y \cdot S , \qquad \quad
			C_{C} =\sum_{i=1}^{N} w_{i} c_{i}^{2}-C \cdot C , \\
			S_{S} & =1-\sum_{i=1}^{N} w_{i} c_{i}^{2}-S \cdot S , \qquad
			C_{S} =\sum_{i=1}^{N} w_{i} c_{i} s_{i}-C \cdot S ,
		\end{split}
	\end{equation}
where $w_i=\left( 1/\sigma^2_i \right) / \left(\sum_{i=1}^{N}1/\sigma^2_i \right)$ is the normalized weight, $s_i=\sin(2\pi \nu t_i)$, $c_i = \cos(2\pi \nu t_i)$, $Y=\sum_{i=1}^{N}w_iy_i$, $C=\sum_{i=1}^{N}w_i \cos(2\pi \nu t_i)$, and $S=\sum_{i=1}^{N}w_i \sin(2\pi \nu t_i)$.
The power spectrum $P_{\rm LS}(\nu)$ is defined as
	\begin{equation}
		P_{\rm LS}(\nu)=\frac{1}{Y_Y \cdot D}\left[S_S \cdot Y_C^{2}+C_C \cdot Y_S^{2}-2 C_S \cdot Y_C \cdot Y_S\right],
	\end{equation}
where $D=C_C S_S - C_S^2$.
Additionally, the significance of the peak values in $P_{\rm LS}(\nu)$ can be assessed by the False Alarm Probability (FAP).
The FAP quantifies the probability of periodic signals arising from random fluctuations \cite{VanderPlas2018}, thereby enabling to exclude false periodic signals.
We implement this analysis using the Python package \emph{\sc Astropy}, but no periodic signals have been verified in the PA shift $\phi (t)$ of FRB 20220912A. 
The results from the time series of polarization measurements for FRB 20220912A with ionospheric corrections are shown in Fig.~\ref{Figure 3}, where the frequency resolution is 0.002 day$^{-1}$.

\hypertarget{MC}{}
\subsubsection*{LS Periodogram-Monte Carlo Method}

To obtain a robust constraint on the ALP-photon coupling constant $g_{a \gamma}$, we reference and adjust the method in Refs. \cite{Ivanov2019, Castillo2022}.
We perform Monte-Carlo simulations to generate the artificial time series that keep the temporal coordinates and exhibit periodic oscillations based on the real distributions of 674 PA data with ionospheric corrections.
This enables us to simulate the power spectrum $P_{\rm LS}(\nu)$ in the presence of the ALP-induced PA oscillations, and to estimate 95\% CL upper limits on $g_{a \gamma}$ by comparing it with $P_{\rm LS}(\nu)$ from real data.
To illustrate our approach, for a quantity $X$, we use distinguishable symbols: $X$ for real data and $\hat{X}$ for simulated data.
For a given frequency $\nu_a$ and the corresponding $P_{\rm LS}(\nu_a)$, the constraint process is summarized as follows:
	\begin{enumerate}
		\item First, we generate 2500 sets of simulated PA data $(\hat{\phi}, \hat{\sigma}, t )$ by randomly sampling from the histogram distributions of the full PA dataset, and insert a periodic signal $\Delta \hat{\phi} = \hat{\alpha} \hat{\varphi} \sin (2\pi \nu_a t + \hat{\delta} )$, where the stochastic fluctuation $\hat{\alpha}$ is sampled from a Rayleigh distribution $f(\hat{\alpha})=\hat{\alpha}\exp{\left( -\hat{\alpha}^2/2 \right)}$, the amplitude $\hat{\varphi}$ is uniformly sampled in the range $[0,\,15]$ degrees, and the phase $\hat{\delta}$ is uniformly sampled in the range $[0,\,2\pi]$.
		\item Next, we calculate $\hat{P}_{\rm LS}(\nu_a)$ for each of the 2500 sets of simulated PA data and pair it with the amplitude $\hat{\varphi}$ to form an array $  (\hat{\varphi}, \hat{P}_{\rm LS}(\nu_a ) )$.
		\item Then, we extract all amplitudes $\hat{\varphi}$ that yield the same power spectrum as the real data at frequency $\nu_a$.
        Specifically, we search for $\hat{\varphi}$ in the array $ (\hat{\varphi}, \hat{P}_{\rm LS}(\nu_a ) )$ that satisfies the condition that the simulated spectrum value $\hat{P}_{\rm LS}(\nu_a)$ falls within a narrow interval $[P_{\rm LS}(\nu_a)-\epsilon, \,P_{\rm LS}(\nu_a)+\epsilon]$ of the real spectrum value. In other words, all $\hat{\varphi}$ must satisfy $\hat{P}_{\rm LS}(\nu_a) \in [P_{\rm LS}(\nu_a)-\epsilon,\, P_{\rm LS}(\nu_a)+\epsilon]$ (here we set $\epsilon=0.005$).
		\item In the set $\hat{\varphi}$ extracted in step 3, we determine a value $\varphi_{95}$ such that 95\% of $\hat{\varphi}$ satisfy $\hat{\varphi}< \varphi_{95}$, representing a 95\% CL upper limit of ALP-induced amplitude at frequency $\nu_a$.
        Finally, an upper limit on $g_{a \gamma}$ for an ALP mass $m_a$ can be inferred with a 95\% CL by providing a SD of $\varphi_{95}/\sqrt{2}$ and a frequency $\nu_a$ according to Equation (\ref{eq:PAstd}).
	\end{enumerate}

\begin{figure*}
\centering
\includegraphics[width=0.9\textwidth]{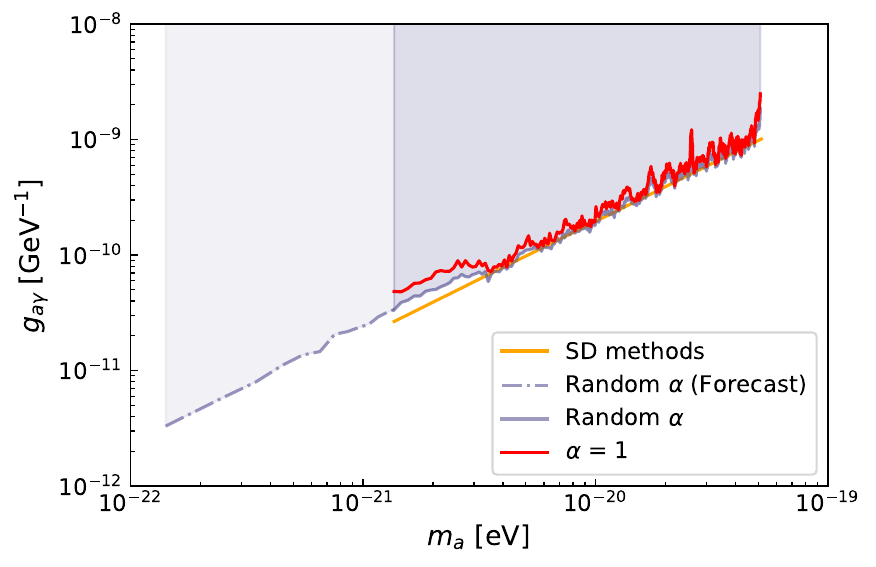}
\caption{
\textbf{The resulting constraints on the ALP-photon coupling constant $g_{a \gamma}$ for different ALP mass $m_a$.}
The orange line represents the estimation from the standard deviation (SD) method.
The purple solid line and red line correspond to the derived upper limits of $g_{a \gamma}$ for the cases of random $\alpha$ and fixed $\alpha=1$, respectively (see Equation (\ref{eq:PA2})).
The purple dot-dashed line represents future constraints from
continued polarization observations of FRB 20220912A up to one year.
}
\label{Figure 5}
\end{figure*}

The above process is performed for each $\nu_a \in \left \{\nu \mid 1/38 \leq \nu \leq 1 \right \}$ (in units of ${\rm day^{-1}}$) to obtain 95\% CL upper limits of $g_{a \gamma}$ for all sensitive ALP mass ranges.
The constraint results (purple shaded area) are presented in Fig.~\ref{Figure 5}, where the results obtained
from the case of fixing $\alpha=1$ (i.e., without considering the stochastic nature of the amplitude of the axion field; 
denoted by red line) and the SD method (denoted by orange line) are also plotted for comparison. 
Similar results indicate that the estimation from the SD method of the daily PA medians is reasonable.
    
The significant variations in PA within a single day are not relevant to the long-period signal we focus on (1–38 days). This is because long-period signals change very little over short timescales, effectively adding a constant to the daily PAs.
Additionally, we can infer that the effects of the stochastic nature of DM are negligible in our analysis,
which can be attributed to the expectation of the Rayleigh distribution being $\sqrt{\pi/2}$, which is close to the fixed value of 1.
We obtain upper limits on the ALP-photon coupling constant of $g_{a \gamma}<(3.4 \times 10^{-11}-1.9\times 10^{-9})\,\mathrm{GeV}^{-1}$ 
for the ALP masses $m_a \sim (1.4\times10^{-21}-5.2\times10^{-20})$ eV.
Based on the current distribution and accuracy of PA data, we simulate the PA variation over the course of a year and apply the same methods to find $\varphi_{95}$.
With future polarization observations of FRB 20220912A for one year, 
the $g_{a \gamma}$ limit would extend to lower ALP masses, i.e., $g_{a \gamma}<3.3 \times 10^{-12}\,\mathrm{GeV}^{-1}$ for an ALP mass $m_a \sim 1.4\times10^{-22}$ eV.

\subsubsection*{The Influence of the Faraday Rotation}

Stable PAs are essential requirements for our study.
For most FRBs, the dominant mechanism of PA variation is the Faraday rotation effect when propagating through magnetized plasma.
The RM is a crucial parameter for quantifying this effect.
The PAs induced by Faraday rotation can be expressed as $\phi = {\rm RM} (c/\nu)^2$, where $c$ is the light speed and $\nu$ is the frequency.
In our FRB 20220912A dataset, the mean value of the daily median RM is $\langle {\rm RM_{Med}} \rangle=-0.91 \pm 1.12 \, {\rm rad\,m^{-2}}$.
Such a small RM may imply that the RM contribution from its host galaxy is comparable to that of the Milky Way \cite{Zhang2023}, which is $\sim -16 \, {\rm rad\,m^{-2}}$ \cite{Hutschenreuter2022}.
Nevertheless, the RM from the galactic medium is stable and thus has negligible influence.
Additionally, the long-term variation of RM is also stable \cite{Zhang2023}, with a slope of $0.017 \,{\rm rad\,m^{-2} \, day^{-1}}$ ($0.06 \, ^{\circ}\,{\rm day^{-1}}$).
The SD of ALP-induced PA shift is $\sim 2^{\circ}$ at a mass of $10^{-21}$ eV, which can be estimated using Equation (\ref{eq:PAstd}). 
If the Faraday rotation contributes a comparable PA shift, the RM needs to reach $\sim 0.6 \, {\rm rad\,m^{-2}}$.
The SD of $\langle {\rm RM_{Med}} \rangle$ for FRB 20220912A is comparable with this value.
This indicates that although it is reasonable to disregard the PA caused by the magnetic field in our analysis, the influence of the Faraday rotation prevents us from further improving accuracy.

\subsubsection*{The Influence of the Circular Polarization Degree}

\begin{figure*}
\centering
\includegraphics[width=0.9\textwidth]{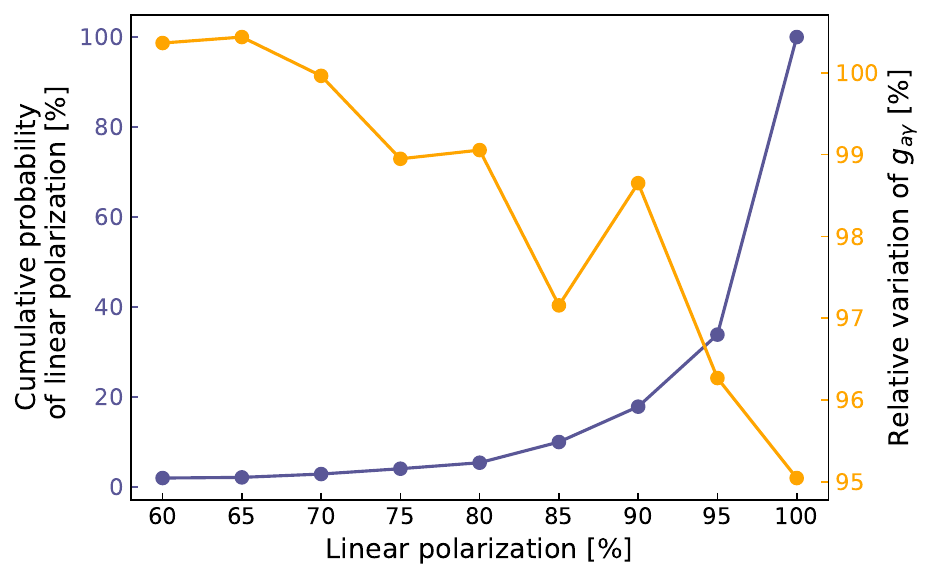}
\caption{
\textbf{The cumulative probability distribution of the linear polarization degrees of 674 bursts from FRB 20220912A and the relative variation of the coupling constant $g_{a \gamma}$.}
The left $y$ axis: the cumulative probability distribution of the linear polarization degrees (purple line).
The right $y$ axis: the relative variation of the coupling constant $g_{a \gamma}$ obtained by the standard deviation (SD) method after removing those data below a given linear polarization threshold and then dividing by the $g_{a \gamma}$ constraint obtained from the total dataset (orange line). 
}
\label{Figure 6}
\end{figure*}

Despite the highly linear polarization observed in FRB 20220912A, observations have shown that this source also exhibits 
a fraction of circular polarization. The possible explanations include intrinsic radiation mechanisms \cite{Wang2022, Qu2024} 
or propagation effects within a magnetar's magnetosphere \cite{Lyutikov2022, Zhao2024}. To investigate the influence of the circular
polarization degree, we calculate the cumulative probability distribution of the linear polarization degree of 674 bursts 
and employ the SD method to quantify constraint results. Fig. \ref{Figure 6} presents the cumulative probability distribution 
of linear polarization, which shows that over 80\% of bursts have a linear polarization fraction exceeding 90\%.
After removing those PA data below a given linear polarization threshold, we can calculate the coupling constant $g_{a \gamma}$ 
for the remaining PA data. This $g_{a \gamma}$ value is then divided by that obtained from the total PA dataset for normalization. 
Taking the linear polarization threshold from 60\% to 100\%, the relative variations of $g_{a \gamma}$ are also depicted 
in Fig. \ref{Figure 6}. The absence of a significant reduction in $g_{a \gamma}$ after removing the low polarization degree data
suggests that non-linearly polarized data have little impact on our results.

\subsubsection*{The Influence of the Dark Matter Density}

\begin{figure*}
\centering
\includegraphics[width=0.9\textwidth]{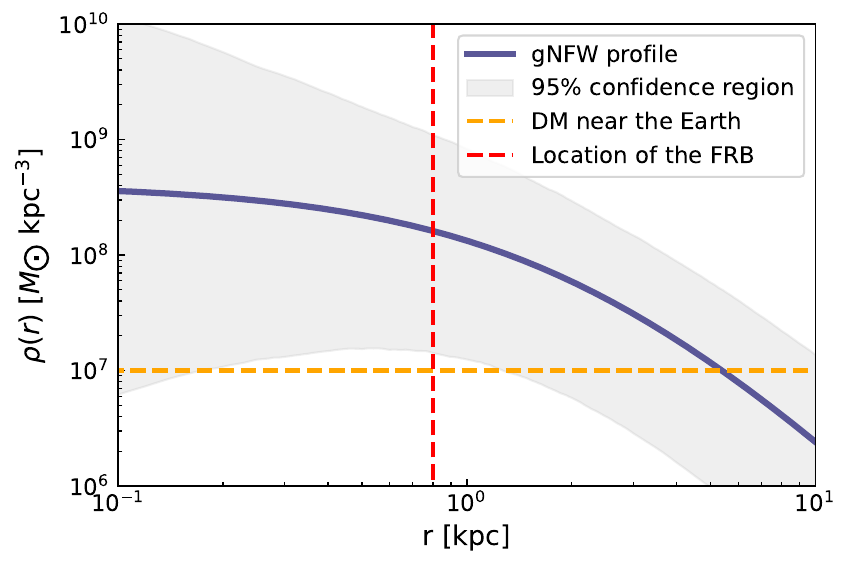}
\caption{
\textbf{The best-fitting generalized Navarro-Frenk-White (gNFW) profile of NGC 4451 (purple solid line) with a 95\% confidence region (gray shade).}
The relation between the dark matter (DM) density $\rho(r)$ and the distance $r$ from the galaxy center is displayed (see Equation~(\ref{eq:darkmatter})).
The location of FRB 20220912A (red dashed line) and the estimated DM density near our Earth based on the Galactic NFW profile (orange dashed line) are indicated.
}
\label{Figure 7}
\end{figure*}

The assumption of DM density employed in this paper is approximate, but it can be shown that its impact on our results is negligible.
The precision of the VLBI localization corresponds to a
physical length of less than 10 pc \cite{hewitt2023}, allowing us to disregard the location error.
The errors associated with the generalized NFW profile, characterized by 
three parameters and their errors, are as follows: $\rho_0=0.41^{+0.47}_{-0.24} \,M_{\odot}\,{\rm pc^{-3}}$, $r_s=2.2^{+0.8}_{-0.7}\, {\rm kpc}$, and $\beta = 0.0^{+0.5}_{-0.6}$ \cite{Cooke2022}.
We estimate the errors in DM density by randomly sampling parameters from their error distributions.
The best-fit generalized NFW profile of NGC 4451 and its 95\% CL are shown in Fig. \ref{Figure 7}.
We find that the 95\% CL lower limit of DM density at a radius of 0.8 kpc is similar to that near the Earth.
Therefore, in the worst-case scenario, the DM density is set to a value near the Earth ($\rho\sim 0.01 \,M_{\odot}\,{\rm pc^{-3}}$), resulting in a coupling constant limit $g_{a \gamma}$ that is four times larger.
In general, since $g_{a \gamma}$ is proportional to the square root of DM density, any estimation deviation in DM density must be significant to meaningfully affect our results.

\clearpage
\section*{References}
\vspace{1cm}

\clearpage

\subsubsection*{Data availability}
Raw data are available from the FAST data center: \url{http://fast.bao.ac.cn}. Owing to the large data volume, we encourage contacting the corresponding author for the raw data transfer. 
The processed PA data and the derived constraints on ALP-photon coupling constant $g_{a \gamma}$ from the FRB polarizations are available in Supplementary Data 1.

\subsubsection*{Code availability}

\noindent \emph{\sc HEIMDALL} (\url{https://sourceforge.net/projects/heimdall-astro})

\noindent \emph{\sc PSRCHIVE} (\url{http://psrchive.sourceforge.net})

\noindent \emph{\sc PAC} (\url{https://psrchive.sourceforge.net/appendices/basis/basis.pdf})

\noindent FRion (\url{https://github.com/CIRADA-Tools/FRion})

\clearpage

\clearpage
\begin{addendum}

\item 
Bao Wang thanks Lei Lei for helpful discussions on dark matter.
This work is partially supported by the National SKA Program of
China (2022SKA0130100), the National Natural Science Foundation of China (grant Nos. 12422307, 12373053, 12321003, and
12041306), the Key Research Program of Frontier Sciences (grant No. ZDBS-LY-7014)
of Chinese Academy of Sciences, International Partnership Program of Chinese Academy of Sciences
for Grand Challenges (114332KYSB20210018), the CAS Project for Young Scientists in Basic Research
(grant No. YSBR-063), the CAS Organizational Scientific Research Platform for National Major
Scientific and Technological Infrastructure: Cosmic Transients with FAST, and the Natural Science
Foundation of Jiangsu Province (grant No. BK20221562).

\item[Author Contributions] J.-J. W. and B. W. initiated the discussion of the project. 
B. W. conducted the bulk of the calculations and drafted the manuscript.
X. Y. and S.-B. Z. processed the polarimetric data of FRB 20220912A detected by the FAST.
J.-J. W. and X.-F. W. discussed the analysis methods and made numerous corrections to the writing.
All authors contributed to the analysis or interpretation of the data and to the final version of the manuscript.

\item[Competing Interests] The authors declare that they have no competing financial interests.

\item[Correspondence] Correspondence and requests for materials should be addressed to Jun-Jie Wei, Song-Bo Zhang and Xue-Feng Wu.

\end{addendum}
\clearpage

\end{document}